\documentclass[prd, aps]{revtex4}

\def\be{\begin{equation}}
\def\te{\end{equation}}
\def\ee{\end{equation}}
\def\ba{\begin{eqnarray}}
\def\bea{\begin{eqnarray}}

\def\tea{\end{eqnarray}}
\def\ea{\end{eqnarray}}
\def\eea{\end{eqnarray}}

\begin{document}

\title {Emergent/Quantum Gravity: \\Macro/Micro Structures of
Spacetime\footnote{\it This is a descriptive summary of the main
themes presented in invited talks in the following international
meetings: Workshop ``From Quantum to Emergent Gravity: Theory and
Phenomenology" Trieste, Italy.  June 11-15, 2007; Peyresq Physics
Meeting, June 16-22, 2007;  Loops '07 Meeting in Morelia, Mexico,
June 25-30, 2007 http://www.matmor.unam.mx/eventos/loops07/; Workshop
on "Condensed Matter meets Gravity", Lorentz Center, University of
Leiden, Holland, August 27-31, 2007; Symposium on Foundations of
Physics, University of Maryland, April 24-27, 2008; Workshop on
"Emergent Gravity", MIT, August 25-29, 2008. Proceedings of DICE08
conference held in Castiglioncello, Italy, Sept.22-26, 2008 will
appear in J. Phys. Conf. Ser. (2009).}}
\author{B.~L. Hu}
\email{blhu@umd.edu}
\affiliation{Department of Physics, University
of Maryland, College Park, Maryland 20742-4111 U.S.A}
\date{March 4, 2009}
\noindent

\begin{abstract}
Emergent gravity views spacetime as an entity emergent from a more
complete theory of interacting fundamental constituents valid at much
finer resolution or higher energies, usually assumed to be above the
Planck energy. In this view general relativity is an effective theory
valid only at long wavelengths and low energies.  We describe the
tasks of emergent gravity from any (`top-down') candidate theory for
the microscopic structure of spacetime (quantum gravity), namely,
identifying the conditions and processes or mechanisms  whereby the
familiar macroscopic spacetime described by general relativity and
matter content described by quantum field theory both emerge with
high probability and reasonable robustness. We point out that this
task may not be so easy as commonly conjured (as implied in the
`theory of everything') because there are emergent phenomena which
cannot simply be  deduced from a given micro-theory. Going in the
opposite direction (`bottom-up') is the task of quantum gravity,
i.e., finding a theory for the microscopic structure of spacetime,
which, in this new view, cannot come from quantizing the metric or
connection forms because they are the collective variables which are
meaningful only for the macroscopic theory (valid below the Planck
energy). This task looks very difficult or almost impossible because
it entails reconstructing lost information. We point out that the
situation may not be so hopeless if we ask the right questions and
have the proper tools for what we want to look for.  We suggest
pathways to move `up' (in energy) from the given macroscopic
conditions of classical gravity and quantum field theory to the
domain closer to the micro-macro interface where spacetime emerged
and  places to look for clues or tell-tale signs at low energy where
one could infer indirectly some salient features of the
micro-structure of spacetime.
\end{abstract}
\maketitle

\newpage
\section{Emergent / Quantum Gravity}

Beneath the diversity of theories and proposals \cite{Oriti} there is
at least general agreement from all camps on the definition of
quantum gravity: It is a theory for the microscopic structure of
spacetime. The major disagreement lies in whether such a theory may
be obtained by quantizing general relativity (GR), a highly
successful theory for the macroscopic structure of spacetime we are
familiar with. The alternative viewpoint is that GR is a low energy
effective theory, and the metric and connection forms are the
collective or hydrodynamic variables of some unknown microscopic
theory. These variables will lose their meaning at shorter
wavelengths and higher energies. Classical gravity in this view is
emergent from quantum gravity (definition above) but not the
classical limit or correspondence of a theory obtained by quantizing
GR. Taking this view the effort in the past half a century in seeking
suitable ways of quantizing this classical theory is considered
largely misdirected, in that doing so will not lead to a microscopic
theory of spacetime. In the analogy of a crystal made of atoms
quantizing the vibrational modes yields phonons, not atoms. Finding
the atomic structure of matter does not come from simply quantizing
its collective degrees of freedom, but takes a very different path.
We will discuss these two routes, namely, finding features in the
microscopic structure of spacetime (quantum gravity) from the given
macroscopic spacetime described  by GR (we shall refer to this as the
`bottom up' approach -- up here refers to energy scale), and
recovering the emergent properties of the familiar macroscopic theory
of spacetime from candidate theories of quantum gravity (`top-down')
by studying the mechanisms and processes of emergence. Thus the title
of this essay.

The emphasis and approaches, the goals and methodology of these two
routes
are very different. To motivate why this new view makes better sense
I have given examples before from atomic - condensed matter physics,
molecular and hydro-dynamics, quantum fluids and critical phenomena
(see below). Instead of finding better ways to quantize GR the new
challenge is to  infer the microscopic structure from the known
macroscopic phenomena, not an easy task as it entails reconstructing
lost information. However, this is not a new demand, it has been the
task for physicists for centuries. For this purpose, concepts and
methods from nonequilibrium statistical mechanics and examples from
strongly correlated many-body systems will probably play an essential
role hitherto largely ignored in quantum gravity.

In the top-down direction one may think that deducing the macro
theory from a \emph{\textit{known}} micro theory is easy. This is the
attitude behind a `theory of everything'.  One can invoke familiar
procedures such as taking the hydrodynamic or thermodynamics limits
in classical physics or using the correspondence principle or
applying effective field theory in quantum physics.  But as we shall
see, this may not be so straightforward as imagined because there are
different types of emergence. If we don't know the underlying
constituents we can learn from the characteristics of many emergent
theories in the description of nature to understand and formulate the
basic rules of emergence. We could then apply the insights gained to
describe the conditions and processes or mechanisms for the emergence
of spacetime. These are the new tasks for emergent gravity in the
`top-down' direction.

\subsection{Emergence}

The view that gravity is emergent can be reached in a number of ways.
My own path of search and discovery is illustrated by the following
observations: a) \textit{Cosmology as `condensed matter' physics}
(1988) \cite{cosCMP}, pointing to the importance of recognizing that
processes like cosmological phase transition, structure formation or
even `birth of the universe' scenarios are not only governed by the
underlying general relativity and quantum field theories, but also
follow the rules more often found in condensed matter physics. b)
\textit{Semiclassical (mean field) and stochastic (fluctuations)
gravity as mesoscopic physics} (1994) \cite{meso}, highlighting the
key role played by noise and fluctuations in many important processes
in cosmology and black hole physics. c) \textit{Decoherence and the
emergence of classical spacetime} from wavefunctions of the universe.
We point out that the outcome of decoherence (see, e.g.,
\cite{decoh}) and the existence of relatively stable quasi-classical
domains resembling our universe depends critically on how
coarse-grainings in the environment or histories are chosen and
implemented, including considerations of the balance between
variability and robustness in the advent of these structures
\cite{timeasy}. And, in an open system \cite{qos} conceptual
framework it is always important to make judicious choices of
collective variables for the best description of macroscopic
phenomena at every level of structure and their interactions (see
also \cite{GidHarMar}). d) \textit{General relativity as
hydrodynamics }(1996) \cite{GRhydro}: conservation laws aids in the
decoherence of long wavelength modes \cite{Halliwell,HartleMarolf}
and the appearance of quasiclassical domains, and  e) \textit{From
stochastic gravity  to quantum gravity} (1999)
\cite{stogra,stograRev}: How to retrieve quantum coherence
information in the gravity sector from metric fluctuations induced by
(higher order) correlations in the quantum field f) \textit{Kinetic
theory approach to quantum gravity} (2002) \cite{kinQG}, via a new
Einstein-Boltzmann hierarchy of equations relating the correlations
in the matter to that in the gravity sector, with the Einstein
equation as the lowest rung of this hierarchy.  g)\textit{Spacetime
as condensate} (2005) \cite{STcond}: Why it is not total nonsense to
view the ultracold spacetime \emph{today} as a quantum entity? What
benefits can one derive from taking this view? What can we learn
about quantum gravity from Bose-Einstein condensate (BEC)? The
overarching goal is to find the microscopic constituents and their
interactions from the collective dynamics of the macroscopic
variables.  (The last three themes will be summarily described in
Sec. 4 below.)

\subsubsection{Emergent Theories Emerging}

In the last three years this viewpoint is increasingly embraced by
researchers from the string theory \cite{Seiberg,HorPol}, particle
and nuclear physics \cite{HorHub,Son,HKSS,Shuryak} communities in the
light of string-gravity-gauge dualities
\cite{AdSCFT,Witten,holography,HorPol}, and pursued by an increasing
number of young researchers in the quantum gravity community as
witnessed in the talks at the present meeting. Two major proponents
of this view before it became popular are Volovik \cite{Volovik} and
Wen \cite{Wen}.

\textit{Emergence is not a strange new concept.} Originated in
biology and sociology but also prevalent in condensed matter physics,
emergence phenomena are more commonly encountered in nature than
constructs deduced from the reductionist paradigm based on linear
logical inductions. An extreme version of the latter is the theory of
everything (TOE) which says that if we knew all the elementary
constituents and their interactions we could deduce the structure and
dynamics of everything in the physical world.

In reality, there are new emergent rules or laws governing the
organization and dynamics of the basic constituents and new modes of
interaction at every level of structure. This counterpoint is
cannonized in Anderson's 1971 outcry (at a time when elementary
particle physics and its philosophy seemed to overshadow other
subfields): ``More is different". Nonetheless it is well-known that
progress in particle physics required the injection of ideas from
condensed matter physics, symmetry breaking being a well-
acknowledged example.

In truth these two aspects are always present in our quest to
comprehend the physical world:  Something assumed to be `elementary'
at a lower energy or with a coarser resolution at one time will later
be found to be a composite at a higher energy or under a finer
resolution. Though their emphasis and approaches are very different
these two paradigms are equally important and functional in the
development of our conception and understanding of nature. The degree
of difficulty may vary. Constructing the macroscopic structure and
dynamics from known microscopic constituents is often easier, such as
getting hydrodynamics from molecular dynamics. Going the other way,
i.e., deducing molecular features from hydrodynamics, is more
difficult. For spacetime structures, the former is the task of
emergent gravity, the latter is the goal of quantum gravity.  Note,
however, that even if we have the precise knowledge of the
fundamental theory, seeing the emergent behavior at lower energies is
not as straightforward as often presumed.

\subsubsection{Deducible and non-deducible Emergence}

To understand the characteristics of emergent theories, and the rules
of emergence one should distinguish in the start those which can be
logically or methodically deduced from a microscopic theory and those
which cannot, at least not without the benefit of knowing certain
attributes of the macroscopic theory. Hydrodynamics from molecular
dynamics or nuclear physics from QCD are examples of the former and
quantum Hall effect is often quoted as an example of the latter: One
can construct useful theories only after such an effect was observed,
but not (or highly unlikely) before. One can appreciate how
nontrivial this task is by looking at the challenges of string theory
or loop quantum gravity in reproducing the physical world of low
energy phenomena.

\subsection{Emergent Gravity: Macro from Micro}

If spacetime is an emergent entity then its large scale structures
should be deductible from the underlying  constituents and their
interactions, like a crystal structure or a condensate constructed
from atoms. Collective (macro) features such as elasticity (in
gravity, metric elasticity \cite{Sak}) associated with such a (micro)
structure  can lend themselves to a nice geometric depiction but
these features will cease to make sense at scales close to or smaller
than an atom. In these cases, the task of emergent gravity is similar
to the derivation of hydrodynamic equations of motion and
thermodynamic laws from molecular dynamics.

\subsubsection{Top-down: Micro to Macro}

Spacetime is the resultant large scale structure formed by the
interaction of sub-level constituent particles. Even when the micro
constituents are believed to be known -- strings \cite{string}, loops
\cite{loop}, spin-foams \cite{spinfoam}, simplicials
\cite{simplicial} --
one still needs to show the macro limit exists and how it is attained
from their interactions. It is a necessary criterion any candidate
theory of quantum gravity needs to meet. For deductive emergent
behavior, path could be tortuous but in principle attainable.

As is known from many examples of condensed matter physics, it is
often a nontrivial task to deduce mesoscopic behavior from
micro-dynamics: one usually encounters nonlinear or even nonlocal
interactions in strongly correlated systems, and needs some ingenuity
to identify collective variables at successive levels of structure to
make the analysis simpler. Using the molecular to hydro dynamics
example, this intermediate scale corresponds to finding the
appropriate kinetic variables and the associated dynamics (e.g., use
of maximal entropy laws at successive stages of complexity).

In addition to nonlinearity, nonlocal properties can emerge from
every level of structure and dynamics. This makes the transition to
macroscopic limit much  more involved. It requires not just a linear
progression of deductions from one single level, but the injection of
new ideas at many levels of construction.

The above assumed that we are dealing with deductive-predictive
theories. But we should be mindful that this macro manifestation
could be via a nondeductive emergent behavior from the micro
structures, in which case even micro- to macro-, sub- to super-
structures would not be easy, some cases could even be impossible.

\subsubsection{Emergence from existing theories of quantum gravity}

We can use some major candidate theories of quantum gravity (micro
structure of spacetime) to examine the key issues listed above. I see
the  causal dynamical triangulation  program pursued by Ambjorn and
Loll et al (see \cite{simplicial} and earlier references therein) as
the more intuitive and operationally accessible option. Regge's
original conception (of Regge Calculus and Regge-Ponsano calculus) is
to reduce the essential ingredients of spacetime to its rudimentary
constructs, i.e., the simplices, such as having the curvature of
spacetime residing on the vertices. One can see how piecing together
these simplices give rise to different spacetime structures, many of
them have fractal dimensions and few grow to the size of our
universe. This scheme is useful to see how the macroscopic structure
emerges, a welcoming recent result is the 4-dimensionality of our
spacetime. However, since the simplices are chosen to contain the key
features of the macroscopic spacetime (e.g., curvature), they are
good building blocks \footnote{ It is like the fine powders ground
from a diamond, even though much smaller in scale, each speckle still
possesses the crystal structure of the macroscopic object. Without a
drastic increase in resolution it cannot reveal the underlying atomic
structure.} yet may contain information about the true degrees of
freedom in the microscopic substructure (like strings or loops, if
they are proven so).  Spin-foam theory \cite{spinfoam} is
aesthetically appealing and rich in mathematical structure. But I
still cannot see how a macro variable such as the connection form
when quantized would turn into a micro-variable (such as the basic
differences between a phonon and an atom). As for string theory, the
AdS/CFT correspondence \cite{AdSCFT,Witten} is often cited as an
example of emergent gravity. This to me is not a real answer. One
needs to show that both gauge theory and gravity emerge from the same
more basic theory, presumably the M theory, but that is still a hope
\cite{FreidelAdS}. To claim emergence each of the candidate theories
of microscopic structure of spacetime needs to identify at least the
principal mechanisms responsible for generating the macroscopic
spacetime with distinct identifiable properties. Gauge-gravity
duality relation \cite{HorPol} has been applied to many branches of
physics from gauge fluid viscosity in RHIC \cite{Son,RHICgrav} to
quantum phase transition in condensed matter (in terms of dyonic
black holes ) \cite{Sachdev} and superconductivity \cite{Herzog} with
surprising results. I will have a short comment on the implications
of these recent developments in the last section.

\subsection{Quest for Quantum Gravity: A theory for the microscopic structures of spacetime}

 Compared to working out the macroscopic limits of a known
 microscopic theory, going the reverse way is always more
difficult because of the missing information. One has only the
macroscopic features at hand which are often grossly coarse-grained
from the microscopic. From the degraded information one hopes to
catch a glimpse of the micro structure. The task is daunting. But
this is how physics has progressed through the centuries.

\subsubsection{Bottom-up: Macro to Micro}

To proceed from bottom up we think that focusing on two key elements
may help:

1) \textit{Topological structures},  because they are more resilient
to evolutionary or environmental changes.  For examples see the works
of Volovik \cite{Volovik}  ($He^3$ analog, Fermi surface) and Wen
\cite{Wen} (string-nets, emergent light and fermions)

2) \textit{Noise-fluctuations}.  Fluctuations could reveal some
sub-structural contents and behavior, such as deducing molecular
interaction from hydrodynamical fluctuations, or finding the
universality classes of systems from the critical exponents in a
phase transition.


\subsubsection{Common features of collective states built from
different constituents}

In the depiction of the structure and properties of matter there are
two almost orthogonal perspectives: One is by way of its constituents
and interactions, the other according to its collective behavior.  If
we regard this chain of QED - QCD - GUT - QG as a vertical
progression depicting the hierarchy of basic constituents, there is
also a horizontal progression in terms of the stochastic -
statistical - kinetic - thermodynamic/hydrodynamic depiction of the
collective states.
There exist similarities between matters in the same collective state
(e.g., hydrodynamics) but made from different constituents.
Macroscopic behavior of electron plasmas are similar in many respects
to the quark-gluon plasma. Indeed, one talks about
magneto-hydrodynamics from Maxwell's theory as well as
magneto-chromo-hydrodynamics from QCD. In this long wavelength,
collision-dominated regime, their behavior can both be adequately
depicted by the laws of hydrodynamics.  The underlying micro-theories
are different, but the hydrodynamic behavior of these constituents
are similar.   The macroscopic, hydrodynamic equations and their
conservation laws are all based on the dynamical and conservation
laws of microphysics (e.g., Newtonian mechanics)  but now in terms of
a new set of collective (hydrodynamic) variables, this new theory of
the macroscopic behavior emerges with a simple yet distinct life of
its own.




\section{Nonlocality and Stochasticity in quantum and statistical mechanics}

With increased attention to quantum information in recent years one
hears more about `nonlocality' in quantum mechanics, referring to the
issues raised in EPR \cite{EPR} and Bell's inequalities. Even though
predictable correlation exists between spatially-separated quantum
states and quantum entanglement can exist between objects outside
each other's light-cones \cite{LinHu2HO}, quantum mechanics is
unambiguously local \cite{UnruhQnl}, as is quantum field theory.
\footnote{ We thus advise against the use of the word `nonlocal' in
this context because it leads to unnecessary confusion.}

Nonlocality in statistical mechanical is an older subject, perhaps
less conspicuous than that which is mis-conjured in quantum
mechanics. We will start with a simple observation which illustrates
two main themes of this exposition: The first theme is that locality
or nonlocality depends sensitively on the level of structure and
dynamics it pertains to or the degree of precision in one's
measurements. Micro-locality in general has nothing to do with
macro-locality. The second theme is that nonlocality (correlations)
and stochasticity (fluctuations) often appear together, and one could
use noise and fluctuations to probe into the nonlocality of
structures and dynamics.

\subsection{Collective Variables constructed from constituent
variables} 

At the risk of appearing naive and pedantic, I'll give two household
examples to illustrate this point.  Perhaps the first collective
behavior we learned in physics is wave phenomena. There, many
individual particles each undergoing a simple harmonic motion in a
well-coordinated manner give rise to waves propagating in the medium
of these particles. Already here we see the relation between the
micro dynamics and the macro behavior. The rules of wave mechanics is
different from that of the constituent particles. The waves obey a
wave equation while the molecules obey Newton's second law. Surely
one can deduce the macro behavior from the micro dynamics but the
reverse relation is many to one: transverse  mechanical and
electromagnetic waves or longitudinal sound waves share the same
waves mechanics description.  One feature to observe relevant to our
theme is the relation of locality to nonlocality. Waves can propagate
from one point in space to another (snapshot graph) which is
correlated with the evolution in time at any one point in space
(history graph). However each individual particle only undergoes a
simple harmonic motion around its point of origin and there is
nothing nonlocal (used here in a non-technical sense) about it.
Another feature to observe is that quantizing the wave degrees of
freedom (sound modes becomes phonon) neither yields the dynamics of
the constituent particles nor reveal any microscopic information
about the molecules. The conception of wave phenomena simply ceases
to exist and the theoretical constructs of wave mechanics irrelevant
at the microscopic level. For the description of atomic motion we
need to use the quantum wave functions for the atom but that is a far
cry from the mechanical waves.

As another commonly encountered example, consider two depictions of
gas or fluid dynamics, one at the level of molecular dynamics, the
other of hydrodynamics. A molecule will collide with many others at
very high speed for a long time before diffusing a short distance.
That molecule's locality in a collective entity as a gas or fluid
element is very different from that induced by following its own
history at the micro scale where multiple collisions take place along
the trajectory. Other molecules arriving at the same time in this gas
or fluid element bear no locality relation with this particular
molecule before or after the instant of collision (assuming a
short-ranged or contact potential).
The smoothness and continuity of the fluid element's movement
described by the Euler equation as reported by a macroscopic observer
belie the stochastic nature of molecular dynamics \footnote{ Quantum
correlations and entanglement have added layers of complexity
\cite{CiracQcorQent}.}.

\subsection{Locality is specific to the choice of observables} 

These two elementary examples show that the notion of locality is
very specific to the  observables chosen for the description of a
particular physical phenomenon of interest (e.g., molecular dynamics
or wave propagation). Nonlocality often appears when one tries to
translate physics expressed in one set of collective variables
suitable for one level of structure to another set.

\subsection{Nonlocality and Stochasticity in statistical mechanics} 

I shall comment here only on temporal nonlocality, i.e., processes
with memory (non-Markov). This is common in open system dynamics
(see, e.g., \cite{HPZ} and references therein). Nonlocality in space
or spacetime is more involved. An example of how this issue is
addressed  can be found in the  causal sets approach of Sorkin
\cite{Sorkin}. For two interacting subsystems the two ordinary
differential equations governing each subsystem can be written as an
integro-differential equation governing one such subsystem, thus
rendering its dynamics non-Markovian, with the memory of the other
subsystem's dynamics carried by the non-local kernels. All this is
happening in a closed system except now one has shifted the attention
to one of its subsystems.  Should the other subsystem possess a  much
greater number of degrees of freedom (called environment) and   are
coarse-grained in some specified way, the nonlocal kernels are then
responsible for the appearance of dissipation and noise. It is the
act of coarse-graining which renders open this particular subsystem
of interest. (For a general discussion, see, e.g.,
\cite{Zwanzig,CH08}.)

Nonlocality is linked to stochasticity in nonequilibrium dynamics.
Nonlocal dissipation and nonlocal fluctuations (colored noise) arise
naturally in the open-system dynamics of Langevin and the
effectively-open system dynamics of Boltzmann, or the dynamics of
correlation hierarchy. These features originate from the choice of
\textit{coarse-graining} measures and \textit{backreaction}
processes. NonMarkovian dynamics is an example of nonlocality in
time, meaning that prior history or past memories enter in the
determination of future time development.

Another angle towards nonlocality and stochasticity is correlation.
Correlation functions measure nonlocality and can be related to
fluctuations. (e.g., one form of fluctuation-dissipation theorem can
be phrased in terms of correlations). Strongly interacting and
correlated systems offer an excellent arena to see how these two
aspects are played out: locality at one level versus nonlocality at a
different level, and how coarse-graining leads to stochasticity. Many
condensed matter physics models offer these insights \cite{KMS}.

\section{Nonlocality and Stochasticity in Quantum Gravity} 

There is probably more confusion in our understanding and use of the
notion of nonlocality in quantum gravity. An example in loop quantum
gravity is how to relate a weave state to spacetime. This was
addressed in a classic paper by Ashtekar, Rovelli and Smolin
\cite{ARS92}: A weave state is a kinematical state designed to match
a given slowly varying classical spatial metric. The concept of
quantum threads (say, from a spin-network) weaved into a fabric
(manifold) of classical spacetime already tacitly assumes (wrongly,
as is now known) a particular kind of micro to macro transition,
where there is a simple correspondence or even equivalence between
locality at the micro and the macro levels. An improvement over the
original suggestion was the so-called nonlocal weave states which
leave marks on the fabrics of spacetime \cite{Bombelli}.  Bombelli,
Corichi and Winker \cite{BomCorWin}  use combinatorial methods to
turn random weave (micro) states into (macro) states of spacetime
manifold. But this remains an oversimplification,  as the low-brow
example of emergence from molecular to hydro- dynamics has
illustrated. Markopoulou and Smolin \cite{MarSmo} recently pointed
out that most weave states including Bombelli et al's assume that
these weave states all satisfy an unstated condition of locality
(edges connect two nodes of Planck length metric distance), and that
there are plenty of weave states which do not satisfy this condition.
The question is:  How does one weave from these nonlocal micro states
a fabric (manifold) with the familiar macroscopic locality?  From our
view, (non) locality at one level may have little to do with (non)
locality at other levels.

To unravel these two aspects in quantum gravity one can first try to
understand how these issues play out in known physical models, then
apply to spacetime, incorporating its special features such as
diffeomorphism invariance (which, in the emergent gravity view, is
associated only with the emergence of spacetime manifold and is thus
no longer an issue at the deeper level of its sub-structures). Many
symmetries we are familiar with associated with the macroscopic
spacetime will become meaningless at the microscopic level (quantum
gravity).   One should also use the various candidate theories of
quantum gravity, such as strings, loops, spin-nets, spin foams and
causal sets, to examine the sense of locality at different levels of
structure,  and see how they could be related to locality in our
macroscopic spacetime.

\subsection{Metric Fluctuations and Stochastic Gravity} 

Keeping gravity classical but allowing for the averaged value of
quantum matter field as source constitutes the theory of
\textit{semiclassical gravity}. Including the consideration of the
fluctuations of quantum matter fields as source constitutes the
theory of stochastic semiclassical gravity (see \cite{stograRev} for
a review). Metric fluctuations, both intrinsic (in the gravitational
field) and induced (by quantum matter fields) are the central figures
in this relatively new theory. With a wider acceptance of the view
that gravity is emergent and that quantizing general relativity does
not yield a theory of the microscopic structure of spacetime,
stochastic gravity's theoretical significance and practical
applications are better recognized and utilized. For instance, the
influence of induced metric fluctuations on cosmological structure
formation is not easily considered in the traditional methods
\cite{RV08}. The Einstein-Langevin (E-L) equation has also been
applied recently to the analysis of backreaction of Hawking radiation
and metric fluctuations near the event horizon of an evaporating
black hole \cite{HuRou07}.

Stochastic gravity can be derived in the quantum open system
conceptual framework \cite{stograRev}. As with all open systems
nonlocality and stochasticity in their dynamics are innate features.
They appear in the nonlocal dissipation and colored noise kernels
which give rise to history-dependent and stochastic trajectories. It
is anticipated that these features are prevalent as one traverses any
two adjacent levels of structure in any of the macro, meso and micro
domains.

\subsection{Stochastic in relation to Semiclassical and Quantum
Gravity} 


Intuitively, the difference between quantum and semiclassical gravity
is that the latter loses all the coherence in the quantum gravity
sector. Stochastic gravity improves on semiclassical gravity in that
partial information related to the coherence in the gravity sector is
preserved. This is reflected in the backreaction from the quantum
fields and manifests as induced metric fluctuations. The noise terms
carry quantum information absent in semiclassical gravity. The
coherence in gravity is related to the coherence in the matter field,
as a complete quantum description should be given by a coherent wave
function of the combined matter and gravity sectors. Since the degree
of coherence can be measured in terms of correlations the strategy of
this approach is to examine the higher correlations of the matter
field, starting with the two point function of the energy momentum
tensor in order to probe into the partial quantum coherence remaining
in the gravity sector. Note that in retrieving this quantum coherence
of the gravity sector, we do not pretend to be constructing a
microscopic theory of spacetime. As long as one uses metric or
connection as a dynamical variable, the theory remains macroscopic.
For our purpose here metric fluctuations can provide precious
information about how spacetime behaves at the interface (phase
transition or cross-over) between the micro and macro domains, not
unlike fluctuations (in, e.g., heat capacity at constant volume) near
a critical point contains valuable information about the universality
class of the system undergoing such a phase transition.

If we view classical gravity as an effective theory, i.e., the metric
or connection functions as collective variables of some fundamental
constituents which make up spacetime  in the large and general
relativity as the hydrodynamic limit, we can also ask if there is a
mid-way weighing station like kinetic theory from molecular dynamics,
from quantum micro-dynamics to classical hydrodynamics. This
transition involves both the micro to macro transition and the
quantum to classical transition, characteristics of the mesoscopic
regime. Stochastic gravity is of the nature of a mesoscopic theory:
the noise in the Einstein-Langevin (E-L) equation contains the 4th
order correlation of the quantum field (or gravitons when considered
as matter source). One can work out higher correlation functions of
the quantum matter field and obtain from a generalized E-L equation
the higher order induced metric fluctuations.

In our research program  several routes have been proposed to move
from the bottom up, that is, starting from the well-established
foundation of semiclassical gravity, adding the considerations of
fluctuations and trying to inch our way towards the macro-micro
interface from below, with the hope of unraveling some microscopic
features of spacetime. We describe a few below: a) a kinetic theory
approach to quantum gravity, b) universal `metric conductance'
fluctuations; and c) spacetime as condensate\footnote{The following
section summarizes the ideas described in Sec 1.1. Readers familiar
with them can skip over to the next section.}.

\section{Routes from bottom up} 

\subsection{Spacetime Correlations and Fluctuations: Einstein-Boltzmann hierarchy of equations}

To better appreciate the relation of this mesoscopic regime between
the macro and the micro, we need to explain the  representation of
kinetic theory in terms of correlation dynamics. In \cite{CH2000}
Calzetta and I introduced the master effective action for an
interacting quantum field where one can obtain the hierarchy of
Schwinger-Dyson equations. From this one can derive the kinetic
equations \cite{CH88} and  an expression for the correlation noise
arising from the slaving of the higher correlation functions. The
resulting equation is known as the stochastic Boltzmann equation.
This kinetic theory structure emerging from an interacting quantum
field illustrates how the macro (hydro) dynamics is linked to the
micro (molecular) dynamics. The minor difference is that there are
two distinct sectors, classical gravity and quantum matter field.
This enables one to adopt an open system approach in treating the
relation between them, such as the Einstein-Langevin equation, while
using the Boltzmann-BBGKY hierarchy to systemize the quantum matter
field sector. The linkage between these two sectors exists at every
level of structure and dynamics, the lowest being the semiclassical
Einstein equation which involves only the expectation value of the
stress energy tensor. The next level up is stochastic gravity which
involves the two point function of the stress-energy tensor. The
hierarchy of equations governing the higher order induced metric
fluctuations from the higher correlations of the stress-energy tensor
of the quantum field is what I called the Einstein-Boltzmann
hierarchy of equations which provides a staircase towards the
micro-structure in this kinetic theory approach to quantum
gravity\footnote{This is not the Einstein-Boltzmann equation in
classical general relativity and kinetic theory -- it frames the
classical matter in the Boltzmann style as source of the Einstein
equation. Our Boltzmann-Einstein hierarchy of equation refers to the
layers of structure in the gravity sector. `Boltzmann' is to show the
kinetic theory nature, and `Einstein' to show its spacetime
structure, albeit these two giants provide their respective theories
only for the lowest correlation order and its dynamics: in
distribution function of spacetime and in geometro-hydrodynamics
respectively.}.

\subsection{Strongly correlated systems: Spacetime Conductance Fluctuations}

In earlier expositions of this subject matter, I proposed to view
semiclassical and stochastic gravity as mesoscopic physics
\cite{meso}. Viewing the issues of correlations and quantum coherence
in the light of mesoscopic physics, the stress-energy two point
function on the right hand side of the Einstein-Langevin equation is
analogous to conductance which is given by the current-current two
point function. What this means is that we are actually calculating
the transport function of (the matter particles as depicted by) the
quantum fields. Following Einstein's keen observation that spacetime
dynamics is determined by (while also dictates) the matter (energy
density), we expect that the transport function represented by the
current correlation in the matter (fluctuations of the energy
density) may also have some physical significance or even (perhaps
less refined) a geometric representation at a higher energy scale
(akin to the pseudo-Riemannian spacetime of GR at the low energy
limit).

This is consistent with viewing general relativity as hydrodynamics:
Conductivity, viscosity and other transport functions are
hydrodynamic quantities. For many practical purposes we don't need to
know the details of the fundamental constituents or their
interactions to establish an adequate depiction of the low or medium
energy physics, but can model them with semi-phenomenological
concepts (like mean free path and collisional cross sections)
\footnote{In the mesoscopic domain the simplest kinetic  model of
transport using these concepts are no longer accurate. One needs to
work  with system-environment models and keep the phase information
of the collective electron wave functions, in so doing extending the
Buttiker-Landauer formula.}. When the interaction among the
constituents gets stronger, effects associated with the higher
correlation functions of the system begin to show up. Studies in
strongly correlated systems are revealing in these regards
\cite{mesobooks,mesotrans,Akkermans05}. For example, fluctuations in
the conductance -- from the 4 point function of the current -- carry
important information such as the sample specific signature and
universality. Although we are not quite in a position, technically
speaking, to calculate the energy momentum 4 point function, thinking
about the problem in this way may open up many interesting conceptual
possibilities, e.g., what does universal conductance fluctuations
mean for spacetime and its underlying constituents? (In the same
vein, I think studies of nonperturbative solutions of gravitational
wave scattering \cite{gravscatt} will also reveal interesting
information about the underlying structure of spacetime beyond the
hydrodynamic regime.) Thus, viewed in the light of mesoscopic
physics, with stochastic gravity we may probe into the higher
correlations of quantum matter and the associated excitations of the
collective modes in geometro-hydrodynamics and the kinetic theory of
spacetime micro-dynamics.

\subsection{Spacetime Condensate: the
Universe as the Ultimate Macroscopic Quantum Phenomenon} 


Working within this novel conceptual framework of
geometro-hydrodynamics, we suggest a new way to look at the nature of
spacetime inspired by Bose-Einstein condensate (BEC) physics. We ask
the question whether spacetime could be considered a condensate, a
low temperature quantum entity, and if so, its immediate
implications. This is an utterly unconventional thought because the
universe below the Planck energy is still at a very high temperature,
and even more radically, it implies macroscopic quantum phenomena
observed in today's universe. The challenge here is to see if such a
view could make outstanding issues like the dark energy easier to
understand. Note also that unlike BEC in atomic physics we want to
describe its salient features even \textit{without the knowledge of
what the `atom of spacetime' is}.
Please refer to \cite{STcond} for background discussions such as the
meaning of a condensate in different context, how far this idea can
sustain, its advantages and pitfalls, and its implications on the
basic tenets of physics and existing programs  of quantum gravity.

\subsubsection{Unconventional view 1: All sub-Planckian physics are
low temperature physics}

We believe that the physical laws governing today's universe are
valid all the way back to the GUT (grand unification theory) and the
Planck epochs, even at the Planck temperature $T_{Pl}= 10^{32}K$.
Since the spacetime structure is supposed to hold for all
sub-Planckian eras, if we consider spacetime as a condensate today,
it should have remained so from the Planck era onwards.

In fact, one should push this concept to its limit and come to the
conclusion that all known physics today, as long as a smooth manifold
structure remains valid for spacetime, the arena where all physical
processes take place, are low-temperature physics. Spacetime
condensate began to take shape at below the Planckian temperature
$T_{Pl}$, according to our current understanding of the physical
laws. In this sense spacetime physics as we know it is low
temperature hydrodynamics, and, in particular, today we are dealing
with ultra-low temperature physics, similar to superfluids and BECs.

The metric or connection forms are hydrodynamic variables, and most
macroscopic gravitational phenomena can be explained as collective
modes and their excitations (of the underlying micro-theory): from
gravitational waves in the weak regime as perturbations, to black
holes in the strong regime as solitons (nonperturbative solutions).
There may even be analogs of turbulence effects in
geometro-hydrodynamics, possibly predictable and detectable when our
numerical techniques and observational skills are improved.

\subsubsection{Unconventional view 2: Spacetime is, after all, a
quantum entity}

An even more severe difficulty in viewing spacetime as a condensate
is to recognize and identify the quantum features in spacetime as it
exists today, not at the Planck time. The conventional view holds
that spacetime is classical below the Planck energy, but quantum
above. That was the rationale for seeking a quantum version of
general relativity, beginning with quantizing the metric function and
the connection forms. Our view is that the universe is fundamentally
a quantum phenomena, but at the mean field level the many body wave
functions (of the micro-constituents, or the `atoms' of spacetime)
which we use to describe its large scale behavior (order parameter
field) obey a classical-like equation, similar to the Gross-Pitaevsky
equation in BEC, which has proven to be surprisingly successful in
capturing the large scale collective dynamics of BEC
\cite{PethickSmith}, until quantum fluctuations and strong
correlation effects enter into the picture \cite{ReyHFB}.

Could it be that the Einstein equation depicting the collective
behavior of the spacetime quantum fluid has the same footing as a
Gross-Pitaevsky equation for BEC? The deeper layer of structure is
ostensibly quantum, it is only at the mean field level that the
many-body wave function is amenable to a classical description
\footnote{In truth, for any quantum system which has bilinear
coupling with its environment or is itself Gaussian exact (or if one
is satisfied with a Gaussian approximation description) the equations
of motion for the expectation values of the quantum observables have
the same form as its classical counterpart. The Ehrenfest theorem
interwoven between the quantum and the classical is one common
example under these conditions.}.

\subsubsection{The Universe as a Macroscopic Quantum Phenomenon}

The obvious challenge is, if the universe is intrinsically quantum
and coherent, where can one expect to see the quantum coherence
phenomena of spacetime? Here again we look to analogs in BEC dynamics
for inspiration, and there are a few useful ones, such as particle
production in the collapse of a BEC (called bosenova in the
experiment of Donley et al \cite{Donley}).  One obvious phenomenon
staring at our face is the vacuum energy of the spacetime condensate,
because if spacetime is a quantum entity, vacuum energy density
exists unabated for our present day late universe,  whereas its
origin is somewhat mysterious for a classical spacetime in the
conventional view. It is highly desirable to explore the implications
of this view on the cosmological constant and coincidence problems.
One encouraging fact is that in most condensed matter systems the
collective excitations have energies of the same order of magnitude
as the ground state, rarely with a severe discrepancy as in particle
physics scales (the gauge hierarchy problem). The observed fact that
the dark energy is 0.7 of the total budget and not $10^{-120}$ came
as a big surprise. But not so much so from the viewpoint of spacetime
as a condensate. The vacuum energy of the condensate and its
excitations are similar in magnitude. One can investigate this issue
from this perspective, using well-studied models of superconductivity
and superfluidity to explore the implications of viewing spacetime as
a condensate \cite{cosconhydro}.

\section{Quantum + Gravity: Look harder at low energy physics} 

Rather than forcing the union of quantum mechanics (QM) and general
relativity (GR), such as is attempted to quantize GR, to try to get
the microscopic structures of spacetime, it pays to explore more
thoroughly the intersection of these two theories at today's low
energy in order to find out the true nature of their conflicts and
their implied meanings. Perhaps both theories lose their meaning and
applicability at a higher energy.   In the emergent gravity view, GR
is an effective theory which is no longer valid at the Planck energy.
There are also speculations that quantum mechanics is also an
emergent theory which fails at a deeper level where nonlocality and
stochasticity appear \cite{GRWP,Diosi,Gisin,Penrose,Adler}.
Our improved understanding of basic issues like quantum measurement
brings forth new issues in this arena such as the effect of gravity
in quantum decoherence \cite{gravdec} and collective effects in
macroscopic quantum systems \cite{mqp}. Penrose has made challenging
theoretical claims \cite{PenroseDec} and proposed interesting
experimental tests \cite{CohMirror} in this realm. We share the same
view in the underlying assumption of Penrose's proposal, namely, low
energy physics could be useful in our search for a new theory which
encompasses both GR and QM as limiting cases of it.
We mention two aspects worthy of further probing and deliberations.

\subsection{Quantum and gravitational low energy phenomena} 

To explore the interplay of quantum mechanics or field theory and
gravitational phenomena at today's energy, one could start with the
simplest set up of the interaction between a particle (with charge
and mass) or atom (entity with internal degrees of freedom) with a
gravitational field under the influence of a quantum field, including
finer considerations of quantum correlation, entanglement and
decoherence. This is necessary for designing laboratory experiments
for  high precision tests of  quantum effects in gravitation and
cosmology. It can also address basic issues of quantum measurement.
These preparatory probes are useful for finding anomalies or `cracks'
in the existing theoretical structure which is the starting point of
the `bottom-up' approach to QG.

In a quantum field a classical particle will not move along a
geodesic \cite{DalMaz} because of {the influence of quantum
fluctuations}. We know how these effects show up as stochastic forces
acting on its mean trajectory, i.e., via the Langevin equation
\cite{JH1,GH1}. Particle motion is affected both by the intrinsic and
induced (called active and passive by Ford \cite{Ford}) fluctuations
of the quantum field. Studying the effects of an {atom in a
gravitational field} (see, e.g., \cite{ParkerAtom}) and the {effects
of a gravitational field on atom-optical or quantum fluid devices}
(e.g., SQUID) can also provide some channel to explore subtle effects
hitherto unknown arising from the interplay between quantum systems
and a classical gravitational field. One could tap into the
capabilities of today's atomic-optical and superconductor /
superfluid experimentations which have reached a high precision level
for these purposes (many experimental schemes have been proposed,
see, e.g. \cite{QtestGR,LATOR}) .  



{From a theoretical viewpoint many of the systems of relevance to our
goals are at the mesoscopic scale. In order to check against the
results from prospective precision experiments our current
understanding of the interaction of mesoscopic quantum systems with
gravity need be improved \cite{Kiefer}. The interaction of strongly
correlated quantum systems with coherent classical gravitational
fields using, e.g., the Schwinger-Keldysh effective action method for
quantum transport \cite{RammerSmith}), real time description of
macroscopic and dissipative quantum tunneling  \cite{Takagi} (which
can be applied to `Birth' of the Universe scenarios), and analysis of
the properties of the stochastic Schrodinger equation in the presence
of gravity are just a few sample problems.

Coherence (even for classical fields) from a much extended space and
time range can affect the local observations. Thus {all nonlocal and
memory effects should be properly accounted for}. (One sees this in
gravitational radiation reaction). {The best way to ensure these
effects are not lost is to do a quantum field calculation from the
beginning and only at the end take the non-relativistic quantum
limits to compare the results with experiments.} The physics we are
probing is very delicate, thus prompting this caution.

These land-based or space experiments in Earth's vicinity are
designed to test the interaction of quantum probes (atom-optical or
superconductivity devices) with weak gravity, but to really get to
the heart of the problem, one needs to probe quantum effects in
strong gravitational fields, such as near a black hole or in the
early universe. Obviously it is not easy to do experiments near a
real black hole but analog models may offer some valuable insights
\cite{analogG}. By contrast, observations in cosmology can provide
bounds on quantum processes in the early universe, at the GUT or even
the Planck scales. Investigations like this could provide some hints
on how macro features may emerge from the underlying constituents.
Again we do not want to imply that such studies whose backdrop
assumes a manifold structure whose dynamics is described by GR will
give us directly the microscopic structure, but they may reveal some
traits of it at the interface. A more direct link is provided by the
duality between strong gravity (black holes) and quantum gauge fields
\cite{HorPol}  in a holographic view , which we mention below with a
comment.

\subsection{Implications of results from gauge-gravity duality} 

In recent years the gauge-gravity duality relation has been expounded
in application to a variety of sub-fields of physics with astounding
results. Examples are viscosity of strongly interacting gauge fluids
obtained from gravitational perturbations from black holes in an AdS
space \cite{HorHub,Son}, gluon-gluon scattering and deconfinement
transition in QCD \cite{SprVol,deconf} related to Hawking-Page
transition \cite{HawPag} in hot Schwarzschild-AdS space, quantum
phase transition \cite{Sachdev} and superconductivity \cite{Herzog}.

What do these unexpected connections suggest, and what essential
elements enter into the duality? AdS-CFT is built on the holography
principle \cite{holography}, i.e., physics in a classical AdS space
in the bulk is related to conformal field theory (CFT) on the
boundary. Thus one could get information on strongly interacting
quantum gauge fields (such as viscosity in a strongly interacting
fluid or deconfinement transition in QCD) by examining the
corresponding gravity effects in AdS space (such as quasi-normal
modes of black holes and the Hawking-Page transition, respectively).

A very attractive aspect of this approach as I see it is that both
parties entering in this duality relation, quantum gauge field and
classical gravitational field, are familiar objects in ordinary low
energy physics.  Although this duality was first discovered in the
context of string theory one does not need to invoke any progenitor
theory at the Planck scale. This is an example of what we mean by
looking more closely at the relations between quantum (here the
strongly interacting gauge fields) and gravity (here the black holes,
which are non-perturbative solutions in GR) at today's low energy.

Another observation I can make related to the main theme discussed
here is that in all the applications of this duality principle we
have seen so far only classical gravity enters, not quantum gravity.
The only entity in gravity which is quantized is the linear
gravitational perturbations (gravitons, or spin-2 particles, acting
in the same role as phonons in solids). There is no need or place for
quantum geometry in this relation which governs macroscopic phenomena
between quantum gauge
fields and classical gravity. \\

\noindent {\bf Acknowledgment}  I thank the organizers of the
meetings and hosts of my visits where some of these ideas were
presented, in reverse chronological order, Thomas Elze, Olaf Dryer,
Fotini Markopoulou, Laurent Friedel, Lee Smolin, Rafael Sorkin,
Jeffrey Bub, Pawel Mazur, Jan Zaanan, Edgard Gunzig, Alejandro
Corichi, Robert Oeckl and Stefano Liberati for their invitations and
hospitality. I am much obliged to many of my collaborators and
colleagues for discussions on the wide range of topics over the
years, notably, Charis Anastopoulos, Esteban Calzetta, Ted Jacobson,
Albert Roura and Enric Verdaguer. I have also been influenced by
exchanges with or reading the papers of T. K. Ng, Grisha Volovik and
X.-G. Wen. This work is supported in part by NSF grants PHY-0601550
and PHY-0801368.


\end{document}